# Transmitter and Precoding Order Optimization for Nonlinear Downlink Beamforming

Thomas Michel and Gerhard Wunder
Fraunhofer German-Sino Mobile Communications Lab, Heinrich-Hertz-Institut
Einstein-Ufer 37, D-10587 Berlin
{michel,wunder}@hhi.fhg.de

*Abstract*— The downlink of a multiple-input multiple output (MIMO) broadcast channel (BC) is considered, where each receiver is equipped with a single antenna and the transmitter performs nonlinear Dirty-Paper Coding (DPC). We present an efficient algorithm that finds the optimum transmit filters and power allocation as well as the optimum precoding order(s) possibly affording time-sharing between individual DPC orders. Subsequently necessary and sufficient conditions for the optimality of an arbitrary precoding order are derived. Based on these we propose a suboptimal algorithm showing excellent performance and having low complexity.

## I. INTRODUCTION

We study the multi-antenna downlink where each receiver has only a single antenna. From an information theoretic perspective, the channel can be modeled as a multiple-input-single-output (MISO) Gaussian broadcast channel (BC). Assuming that the transmitter has knowledge of the channel and can perform Dirty-Paper Coding, we are interested in the optimum transmit filters, power allocation as well as the *optimum precoding order* to achieve a certain set of SINR targets or equivalently a set of target rates specified in advance. This perspective reflects the recent need for fairness in mobile communication systems and conflicts with throughput maximization.

The problem of sum power minimization has been well studied in the beamforming literature and relying on Perron-Frobenius theory it was solved for linear processing (see e.g. [1], [2] and references therein). Further it was generalized to the case of nonlinear DPC beamforming in [3]. In contrast to the case of purely linear beamforming the system is not interference limited for DPC. However, a drawback of the mentioned approach is that it solves the problem for a *fixed DPC order*. An inherent problem is that for every DPC order the interference coupling matrix has a different structure. So no method is known to jointly optimize the DPC order with the power and transmit filter optimization. Thus finding the optimum DPC order remains a combinatorial problem being prohibitive even for moderate numbers of users.

In order to overcome these problems a generic approach stemming from an information-theoretic perspective con be chosen considering rate constraints instead of SINR constraints (in principle, any QoS-measure related to rate via a bijective mapping can be considered). This allows to use recent results on duality of MIMO BC and multiple access channel (MAC) [4], [5] and sum power minimization for MIMO BC [6]–[9]. Using the duality of MISO BC and single-input-multiple-output (SIMO) MAC the problem was considered for a fixed precoding order in [9] and the same authors showed recently that the precoding order can indeed by optimized using a sequence of convex programs [10].

In this paper we extend the results from [10]. We present an efficient algorithm yielding the optimal power allocation, transmit filters and precoding order exploiting ideas from [11] and [12]. In contrast to the algorithm proposed [10] there is no need to combine two nested loops of convex optimization problems. We study the cases where a given SINR target is achievable with minimum power using time-sharing between different precoding orders pointing out that the information-theoretic approach can be considered as a relaxation. Further we derive necessary and sufficient condition for the optimality of a given DPC. These are easy to check since they rely on a simple recursive equation. Based on these conditions we propose a low complexity algorithm showing very good performance.

The remainder of this paper is organized as follows. Section II briefly describes the system model and Section III contains the problem statement. Subsequently, the problem is solved using a relaxation approach in Section IV. Conditions on the optimality of a DPC order are derived and a suboptimal algorithm is presented in Section V. Numerical results are presented in Section VI and we conclude with Section VII.

### A. Notation

Sets are given by calligraphic letters, lower case bold letters represent vectors and upper case bold letters denote matrices. All vectors are column vectors and $|\cdot|$ is the determinant. A complex variable $c = a + jb$ is said to be circular symmetric complex Gaussian distributed $c \sim \mathcal{CN}(0,1)$ if its real and imaginary part are independently distributed with $a \sim \mathcal{N}(0,1/2)$ and $b \sim \mathcal{N}(0,1/2)$.

## II. SYSTEM MODEL

Consider a MISO downlink, where the transmitter has $n_T$ antennas and each of the $M$ receivers is equipped with a single antenna. Then the signal received by user $m$ can be written as

$$r_m = \mathbf{h}_m^H \mathbf{t} + z_m \qquad (1)$$

*The authors are supported in part by the *Bundesministerium für Bildung und Forschung (BMBF)* under grant FK 01 BU 350

where $\mathbf{h}_m \in \mathbb{C}^{n_T \times 1}$ is the channel matrix to user $m$, $\mathbf{t} \in \mathbb{C}^{n_T \times 1}$ is the transmitted symbol vector and $z_m \sim \mathcal{CN}(0,1)$ is circular symmetric additive white Gaussian noise (AWGN). Assuming that the transmitter has perfect channel knowledge, DPC can be applied. Let $\pi(\cdot)$ be a DPC encoding order such that user $\pi(M)$ is encoded first, followed by user $\pi(M-1)$ and so on. Then the interference caused by users which have been already encoded ($n > m$) can be taken into account while encoding the signal of user $m$ rendering it harmless.

The dual SIMO MAC is described by the system equation

$$\mathbf{y} = \sum_{m=1}^{M} \mathbf{h}_m x_m + \mathbf{n} \quad (2)$$

where $\mathbf{y} \in \mathbb{C}^{n_T \times 1}$ is the received signal, $x_m$ is the symbol transmitted by user $m$ and $\mathbf{n} \in \mathbb{C}^{n_T \times 1}$ is again circular symmetric AWGN. Assume that the receiver can perform Successive Interference Cancellation (SIC). Choosing a decoding order $\pi(\cdot)$ such that user $\pi(1)$ is decoded first followed by user $\pi(2)$ and so on, the signals of already decoded users $n < m$ can be subtracted such that they do not cause any interference while decoding user $m$th signal.

Recent uplink-downlink duality results state that the capacity regions of MIMO BC and its dual MIMO MAC under a common sum power constraint coincide [5] and each rate tuple achievable in the MIMO BC under a certain DPC order $\pi(\cdot)$ can be achieved in the MIMO MAC applying the reverse SIC order [4]. Exploiting these facts all problems can be solved equivalently in the dual channel. Since the SIMO MAC has a much more favorable (polymatroid) structure, we will focus on the uplink to solve the stated problems.

## III. PROBLEM STATEMENT

The problem to be studied can be formulated in a signal-processing context as well as in an information theoretic context. We begin with the signal processing formulation. To this end consider the BC system model in (1) and decompose the transmitted signal to

$$\mathbf{t} = \mathbf{U}\mathbf{s} \quad (3)$$

where $\mathbf{U} = [\mathbf{u}_1, ..., \mathbf{u}_M] \in \mathbb{C}^{n_T \times M}$ is a matrix of spatial beamformers and $\mathbf{s} \in \mathbb{C}^{M \times 1}$ is the vector of independent transmit symbols. Denoting the transmit power used for the symbol of user $m$ with $\tilde{p}_m = \mathbb{E}\{|s_m|^2\}$, the problem can be written

$$\min_{\mathbf{U}, \tilde{\mathbf{p}}, \pi} \sum_{m=1}^{M} p_m \quad (4)$$
$$\text{subj. to } \gamma_{\pi(m)} \geq \bar{\gamma}_{\pi(m)} \quad \forall\, m$$

where $\mathbf{U} = [\mathbf{u}_1, ..., \mathbf{u}_M]^T$ is a matrix consisting of beamformers, $\pi(\cdot)$ is a DPC order such that user $\pi(M)$ is encoded first followed by user $\pi(M-1)$, $\bar{\gamma}_m$ is the SINR target for user $m$ and

$$\gamma_{\pi(m)} = \frac{\tilde{p}_{\pi(m)} \|\mathbf{u}^H_{\pi(m)} \mathbf{h}_{\pi(m)}\|^2}{1 + \sum_{n=1}^{m-1} \tilde{p}_{\pi(n)} \|\mathbf{u}^H_{\pi(n)} \mathbf{h}_{\pi(m)}\|^2} \quad (5)$$

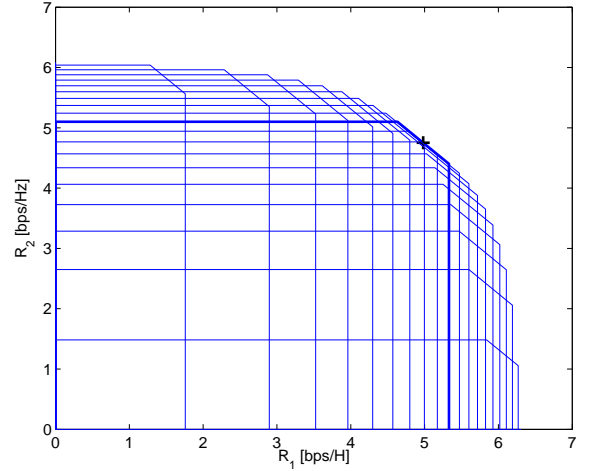

Fig. 1. Exemplary MISO BC (SIMO MAC) capacity region for random 2 user channel and 10 $dB$ transmit SNR. Sum capacity polymatroid highlighted with rate (black marker) tuple *not* achievable with a single DPC order.

This problem has been studied extensively in the signal-processing community (see e.g. [3] and references therein) and can be solved for any fixed precoding order $\pi(\cdot)$ using Perron-root techniques and SINR uplink-downlink duality.

On the other hand (4) can be reformulated in an information-theoretic context if Gaussian inputs are assumed. Introducing covariance matrices $\mathbf{Q}_m = \mathbb{E}\{\tilde{\mathbf{t}}_m \tilde{\mathbf{t}}_m^H\}$ with $\sum_{m=1}^{M} \tilde{\mathbf{t}}_m = \mathbf{t}$ and relating SINR to rate using the bijective mapping

$$R_m = \log(1 + \text{SINR}_m) \quad (6)$$

yields

$$\min_{\mathbf{Q}, \pi} \sum_{m=1}^{M} \text{tr}(\mathbf{Q}_m) \quad (7)$$
$$\text{subj. to } R_{\pi(m)} \geq \bar{R}_{\pi(m)} \quad \forall\, m$$

where

$$R_{\pi(m)} = \log \frac{1 + \sum_{n=m}^{M} \mathbf{h}^H_{\pi(n)} \mathbf{Q}_{\pi(n)} \mathbf{h}_{\pi(n)}}{1 + \sum_{n=m+1}^{M} \mathbf{h}^H_{\pi(n)} \mathbf{Q}_{\pi(n)} \mathbf{h}_{\pi(n)}} \quad (8)$$

and the rate constraints are given by $\bar{R}_m = \log(1 + \bar{\gamma}_m)$.

In order to make (7) tractable continuous relaxation can be used allowing convex combinations of different DPC orders (time-sharing). Combined with duality relations from [4] and introducing uplink powers $\mathbf{p} = [p_1, .., p_M]$ (7) can be reformulated as a convex program in the uplink [10]:

$$\min \sum_{m=1}^{M} p_m$$
$$\text{subj. to } \log \left| \mathbf{I} + \sum_{m \in \mathcal{S}} p_m \mathbf{h}_m \mathbf{h}_m^H \right| \geq \sum_{m \in \mathcal{S}} \bar{R}_m \quad (9)$$
$$\forall \mathcal{S} \subseteq \{1, ..., M\}.$$

**Algorithm 1** Minimum Sum Power for Beamforming

---

(0) initialize $\boldsymbol{\lambda}^{(0)} = [\lambda_1^{max}/2, ..., \lambda_M^{max}/2]$
**while** desired accuracy not reached **do**
  (1) for given $\boldsymbol{\lambda}^{(n)}$ solve

$$\mathbf{R}^{(n)} = \arg\max_{\mathbf{R}} \quad \sum_{m=1}^{M} \lambda_m^{(n)} R_m - \sum_m p_m$$

$$\text{subj. to} \quad R_{\pi(m)} \leq \log \frac{\left|\mathbf{I} + \sum_{n=m}^{M} p_{\pi(n)} \mathbf{h}_{\pi(n)} \mathbf{h}_{\pi(n)}^H\right|}{\left|\mathbf{I} + \sum_{n=m+1}^{M} p_{\pi(n)} \mathbf{h}_{\pi(n)} \mathbf{h}_{\pi(n)}^H\right|} \quad (10)$$

  with DPC order $\pi(\cdot)$ such that

$$\lambda_{\pi(1)}^{(n)} \leq ... \leq \lambda_{\pi(M)}^{(n)}$$

  (2) determine a subgradient $\boldsymbol{\nu} \in \mathbb{R}_+^M$ according to

$$\nu_m = R_m^{(n)} - \bar{R}_m \quad \forall \, m \quad (11)$$

  (3) determine new ellipse with new centroid $\boldsymbol{\lambda}^{(n+1)}$ according to standard ellipsoid algorithm
**end while**

---

The set of constraints corresponding to all partial sums defines the SIMO MAC capacity region. It is important to note that (9) is a *convex relaxation* of the original problems (4) and (7) including time-sharing solutions. This is illustrated Fig. 1, where an exemplary MISO capacity region is depicted. The marked rate tuple is not achievable with a single DPC order at 10 dB although it is a solution to (9).

## IV. A CONVEX RELAXATION APPROACH

The formulation (9) including all rate tuples within the capacity region constitutes the classical sum power minimization problem. It was studied for OFDM (a diagonal MIMO system) in [13] and solved for the general MIMO case in [6], [7], [14]. In principle, (9) is a convex problem and can be solved using standard solvers. However, the number of constraints grows exponentially with $M$ making an optimization quite complex.

In [10] an algorithm comprising two nested loops of convex problems was proposed. The complexity can be considerably reduced using an optimization of the dual function

$$g(\boldsymbol{\lambda}) = \inf_{\mathbf{p}: p_m > 0} \mathcal{L}(\mathbf{p}, \boldsymbol{\lambda}) \quad (12)$$

where $\boldsymbol{\lambda} = [\lambda_1, ..., \lambda_M]^T$ denotes Lagrangian multipliers corresponding to the rate constraints and the Lagrangian is given by

$$\mathcal{L}(\mathbf{p}, \boldsymbol{\lambda}) = \sum_{m=1}^{M} p_m - \sum_{m=1}^{M} \lambda_m (\bar{R}_m - R_m)$$

and $R_m$ is defined in (8). Since the dual (12) is not necessarily differentiable, it can be optimized using any optimization method relying on subgradients such as the ellipsoid method.

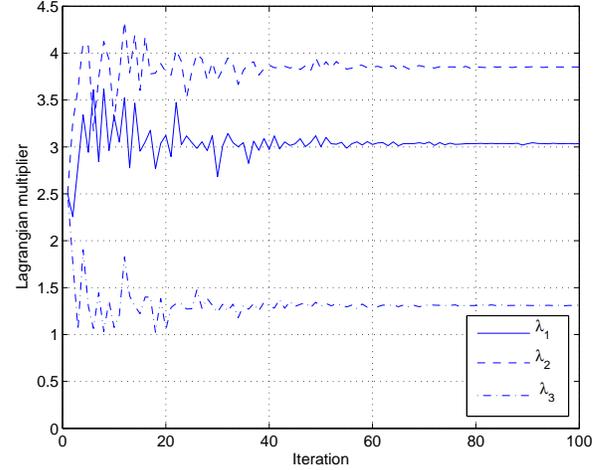

Fig. 2. Exemplary convergence of Algorithm 1 for a channel with $n_T = 3$ and $M = 3$ users.

Further there exists an upper bound on the dual parameters $\boldsymbol{\lambda}$ and thus an initial ellipse can be found (see [7]) [1]. An important observation is that the optimization in (10) can be done very efficiently with a modified version of the algorithm presented in [11]. Although developed for a fixed power budget, it can be modified to solve the problem for fixed $\boldsymbol{\lambda}$ due to a simple monotonicity property (details omitted due to the lack of space). The convergence process is illustrated exemplarily in Fig. 3.

## V. FIXED DPC ORDER AND SUBOPTIMAL ALGORITHM

### A. Fixed ordering

Assume an arbitrary DPC order $\pi(\cdot)$. For this case the original problem (7) simplifies to

$$\min \sum_{m=1}^{M} p_m$$

$$\text{subj. to} \quad \log \frac{\left|\mathbf{I} + \sum_{i=m}^{M} p_{\pi(i)} \mathbf{h}_{\pi(i)} \mathbf{h}_{\pi(i)}^H\right|}{\left|\mathbf{I} + \sum_{i=m+1}^{M} p_{\pi(i)} \mathbf{h}_{\pi(i)} \mathbf{h}_{\pi(i)}^H\right|} \geq \bar{R}_m \quad (13)$$

$$\forall m \in \mathcal{M}.$$

In contrast to (7), there exists a closed-form solution having a recursive structure [9]:

$$p_{\pi(m)} = \frac{2^{\bar{R}_m} - 1}{\left\| \mathbf{h}_{\pi(m)}^H \left(\mathbf{I} + \sum_{i=m+1}^{M} p_{\pi(i)} \mathbf{h}_{\pi(i)} \mathbf{h}_{\pi(i)}^H\right)^{\frac{1}{2}} \right\|_2^2} \quad (14)$$

### B. Conditions for optimality of $\pi(\cdot)$

A question of considerable interest is whether a given precoding order $\pi(\cdot)$ is optimal or not.

---

[1] For details concerning the ellipsoid method the interested reader is referred to [15].

To this end consider the global problem (9). If a precoding order $\pi(\cdot)$ and the corresponding solution $\hat{\mathbf{p}} = [\hat{p}_1, ..., \hat{p}_M]^T$ is optimal, the KKT conditions of (9) have to be fulfilled since (9) is a convex program.

First note that the number of active constraints in (9) is upper bounded by $M$ due to the polymatroid structure. To be more precise define the set

$$\mathcal{S}^K = \{\mathcal{S} \subseteq \mathcal{M} : |\mathcal{S}| = K, K \leq M\}$$

where $|\cdot|$ is the cardinality of a set. Then the number of active constraints $\mathcal{S}_{act}^K \subseteq \mathcal{S}^K$ is at most one:

$$|\mathcal{S}_{act}^K| \in \{0, 1\} \quad \forall K$$

Time-sharing is needed to achieve a certain rate tuple if

$$\exists K : |\mathcal{S}_{act}^K| = 0, \quad (15)$$

i.e. for at least one $K$ there is no constraint active.

In the following we assume that the minimum sum power is attained *without time-sharing*. Then the following Lemma states necessary and sufficient conditions on the optimality of a DPC order:

*Lemma 1:* Let $\hat{\mathbf{p}} = [\hat{p}_1, ..., \hat{p}_M]^T$ be the solution to (13) with DPC order $\pi(\cdot)$. Then $\pi(\cdot)$ and $\hat{\mathbf{p}}$ are the optimal precoding order and power allocation if and only if

$$\lambda_m - \lambda_{m-1} > 0 \quad (16)$$

for all $m = 2, ..., M$ where

$$\lambda_m = \frac{1 - \mathbf{h}_{\pi(m)}^H \left( \sum_{n=1}^{m-1} \lambda_n (Z_n^{-1} - Z_{n+1}^{-1}) \right) \mathbf{h}_{\pi(m)}}{\left\| \mathbf{h}_{\pi(m)}^H Z_m^{-1/2} \right\|^2} \quad (17)$$

and

$$Z_n = \mathbf{I} + \sum_{\substack{i=n}}^{M} \hat{p}_{\pi(i)} \mathbf{h}_{\pi(i)} \mathbf{h}_{\pi(i)}^H \quad (18)$$

*Proof:* Consider the Lagrangian of (13):

$$\mathcal{L}(\mathbf{p}, \boldsymbol{\lambda}, \boldsymbol{\nu}) = \sum_{m=1}^{M} p_m$$
$$+ \sum_{m=1}^{M} \lambda_m \left( \bar{R}_{\pi(m)} - \left( \log \frac{|Z_m|}{|Z_{m+1}|} \right) \right) \quad (19)$$
$$- \sum_{m=1}^{M} \nu_m p_m$$

with $Z_m$ defined in (18). Due to the non-convexity of (13) the KKT conditions are necessary but not sufficient. The derivative w.r.t. $p_{\pi(m)}$ is given by

$$\frac{\partial \mathcal{L}(\mathbf{p}, \boldsymbol{\lambda}, \boldsymbol{\nu})}{\partial p_{\pi(m)}} = 0 = 1 - \nu_m$$
$$- \mathbf{h}_{\pi(m)}^H \left( \sum_{n=1}^{m-1} \lambda_n (Z_n^{-1} - Z_{n+1}^{-1}) \right) \mathbf{h}_{\pi(m)}$$
$$- \lambda_m \mathbf{h}_{\pi(m)}^H Z_m^{-1} \mathbf{h}_{\pi(m)}.$$

Solving for $\lambda_m$ and setting $\nu_m = 0$ (since $p_m > 0$) yields the recursive expression in (17). On the other if $\pi(\cdot)$ is the optimal DPC order, the KKT conditions of the convex global problem (9) must be fulfilled and exactly the $M$ constraints corresponding to the sets

$$\mathcal{S}_m = \{\pi(1), ..., \pi(m)\} \quad m = 1, ..., M$$

must be active. A simple expansion of the Lagrangian given in (19) leads to

$$\mathcal{L}(\mathbf{p}, \boldsymbol{\lambda}, \boldsymbol{\nu}) = \sum_{m=1}^{M} p_m$$
$$+ \lambda_1 \left( \sum_{n=1}^{M} \bar{R}_{\pi(n)} - \log |Z_1| \right)$$
$$+ \sum_{m=2}^{M} (\lambda_m - \lambda_{m-1}) \left( \sum_{n=m}^{M} \bar{R}_{\pi(n)} - \log |Z_m| \right)$$
$$+ \sum_{m=1}^{M} \nu_m p_m$$

which is exactly the Lagrangian for the global problem if $\pi(\cdot)$ is the optimal DPC order and where the Lagrangian multipliers are given by

$$\theta_m = \begin{cases} \lambda_m & m = 1 \\ \lambda_m - \lambda_{m-1} & m = 2, ..., M \end{cases}$$

The KKT conditions of (9) are fulfilled with all constraints specified in (20) being active if and only if

$$\theta_m > 0 \quad \text{for all } m$$

leading to (16). ∎

*Corollary 1:* A DPC order is part of the time-sharing solution if

$$\lambda_{m-1} - \lambda_m \geq 0$$

for all $m = 2, ..., M$ and

$$\exists m : \lambda_{m-1} - \lambda_m = 0$$

for at least one $m$. This corresponds to (15).

### C. A low complexity approximation

In order to develop a low complexity algorithm we exploit the fact that there exists an analytic solution for a fixed DPC order and the optimality condition in (16) is very easy to check due to their simple recursive structure.

An idea based on the optimality conditions is to resort the users according to their Lagrangian multipliers, since they reveal the current *rate price*. Since a user encoded earlier sees less interference he needs less power to support a given rate. Thus it makes sense to encode users with a high value of $\lambda_m$ earlier. Motivated by this argument, Algorithm 2 sorts the users according to their Lagrangian multipliers. By doing so, different vertices are visited. If the optimality condition in (16) is fulfilled or a vertex is revisited the algorithm stops.

**Algorithm 2** Suboptimal Iterative Algorithm

**(0)** fix an arbitrary precoding order $\pi^{(n)}$
**while** not converged **do**
  **(1)** for given $\pi^{(n)}$ find $\hat{\mathbf{p}}(\pi^{(n)})$ by solving minimum sum power problem in (13)
  **(2)** determine new precoding order $\pi^{(n+1)}$ such that
  $$\lambda^{(n)}_{\pi^{(n+1)}(1)} \leq ... \leq \lambda^{(n)}_{\pi^{(n+1)}(M)} \qquad (20)$$
  where $\lambda^{(n)}_m$ is given by (17).
**end while**

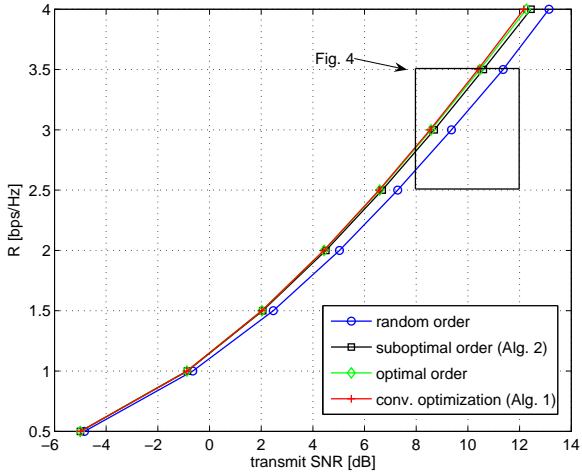

Fig. 3. Rate requirement [bps/Hz] over average transmit SNR [dB] for $n_T = 3$ transmit antennas and $M = 3$ users.

## VI. SIMULATIONS

Figures 3 and 4 illustrate the average transmit power needed to support an equal rate for each of three users, where the transmitter has $n_T = 3$ antennas. Each channel realization was generated according to an iid Rayleigh fading process. It can be clearly observed that the exhaustive search over all DPC orders does not achieve the curve based on convex optimization. This is due to the cases involving time-sharing. The suboptimal algorithm performs surprisingly well despite its low complexity. All three outperform the randomly chose DPC order. Simulations suggest that this gap grows as the number of users increases.

## VII. CONCLUSIONS

We studied the problem of finding the optimal resource allocation and DPC order for nonlinear downlink beamforming, if SINR or rate constraints have to be met. An efficient algorithm based on convex relaxation was presented, which may yield time-sharing solutions. Moreover conditions for the optimality of a specific DPC order were derived and based on these we proposed a low complexity algorithm showing a minimal performance loss.

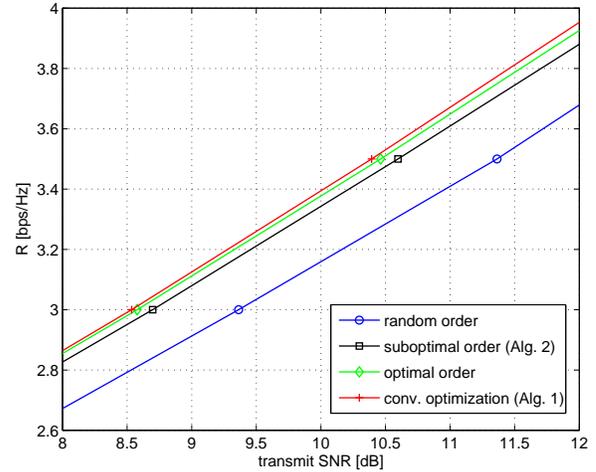

Fig. 4. Enlargement of region marked in Fig. 3.